\documentclass[10pt,twocolumn,letterpaper]{IEEEtran}
\usepackage[ left=1in,
                    right=1.05in,
                    top = 1 in,
                    bottom=1.05in]{geometry}

\usepackage{fancyhdr}

\usepackage{amsmath,epsfig}
\usepackage{algorithm}
\usepackage{bm}
\usepackage{tzplot}
\usepackage[usenames,dvipsnames]{pstricks}
\usepackage{epsfig}
\usepackage{pst-grad} 
\usepackage{pst-plot} 
\usepackage{subfigure}
\usepackage[capitalize]{cleveref}

\usepackage{subfigure}
\usepackage[T1]{fontenc}
\usepackage{amsmath}
\usepackage{graphics}
\include{psfig}
\usepackage{epsfig}
\usepackage{array}
\usepackage{psfrag}
\usepackage{amssymb}
\usepackage{color}
\usepackage{pstricks,pst-node,pst-text,pst-3d,pst-plot}
\usepackage{cite}
\usepackage{ulem}
\definecolor{dgreen}{rgb}{0, 0.8, 0.1}

\usepackage{amsmath}
\usepackage{breqn}
\begin{document}

\title{Entropy-Based Doppler Centroid Estimation and Speckle Noise Reduction for Spaceborne SAR Imaging\\
\author{\IEEEauthorblockN{Shahrokh Hamidi}\\
\IEEEauthorblockA{{Department of Electrical and Computer Engineering}, 
{University of Waterloo}\\
Waterloo, Ontario, Canada \\
shahrokh.hamidi@uwaterloo.ca}
}
}

\maketitle



\thispagestyle{empty}
\pagestyle{plain} 

\begin{tikzpicture}[remember picture, overlay]
      \node[font=\small] at ([yshift=-1cm]current page.north)  {This paper has been published in the 2025 International Conference on Communication and Signal Processing (ICCSP) \copyright IEEE};
\end{tikzpicture}

\begin{abstract}
In this paper, we present strip-map mode spaceborne Synthetic Aperture Radar (SAR) imaging with the focus on Doppler centroid frequency estimation.
The non-zero Doppler centroid frequency is the result of non-zero squint angle which if it is not compensated it can de-focus the image. We present an efficient method based on the entropy of the reconstructed image to estimate the fractional part of the Doppler centroid frequency.  
Furthermore, we discuss the speckle noise, which degrades the quality of the reconstructed images considerably, and attempt to alleviate its effect efficiently. Following the implementation of the speckle noise reduction algorithm, a significant improvement in the quality of the reconstructed images is achieved.
Finally, we utilize the experimental data gathered from the RADARSAT-1 satellite of Vancouver Canada to verify the accuracy and effectiveness of the proposed techniques.
\end{abstract}


\section{Introduction}
Synthetic Aperture Radar (SAR) imaging is a unique method to create the effect of a large real aperture synthetically. The high resolution in the azimuth direction in SAR imaging stems from the relative motion between the radar and the target \cite{Curlander, Soumekh, Cumming, Sullivan}.
In this paper, we specifically consider the strip-map mode spaceborne SAR imaging in which the angle between the main axis of the antenna and the terrain to be imaged, remains fixed \cite{Cumming, Soumekh, Munson_Stripmap}. Similar to all high resolution radars, in SAR imagery systems, the fine resolution in the range direction is achieved by utilizing the idea of pulse compression.
A Linear Frequency Modulated (LFM) signal, also known as chirp signal, is transmitted and the reflected signal is processed using matched filtering technique and as a result a high resolution profile in the range direction is produced \cite{Cumming, Soumekh, Skolnik, Mahafza}. The range resolution depends on the bandwidth of the chirp signal.
One of the features of the strip-map mode SAR imaging is the squint angle of the antenna which shifts the spectrum of the signal in the azimuth direction. For this reason, the signal in the azimuth direction is not at base-band. In other words, the average value of the azimuth frequency, which is called Doppler centroid frequency, is a non-zeros value.
Non-zero squint angle is the main reason for Range Cell Migration (RCM) phenomenon which if it is not compensated, it degrades the image quality in both range and azimuth directions considerably \cite{Cumming}.
The best possible way to estimate the Doppler centroid frequency is by using the received data. Estimating the Doppler centroid frequency by analyzing the geometry does not give rise to an accurate result since the location of the satellite can not be estimated with high precision \cite{Cumming}.
Since the sampling rate in the azimuth direction is equal to the Pulse Repetition Frequency (PRF), therefore, if the Doppler centroid frequency exceeds the PRF, there will be ambiguity in its estimation. In fact there will be two parts for the Doppler centroid, namely, fractional part and the integer part. 

In this paper, we present spaceborne SAR image reconstruction operating in strip-map mode with the focus on estimating the fractional part of the Doppler centroid frequency and speckle noise removal. The techniques which have been proposed for the Doppler centroid frequency estimation in the literature are either phase-based or amplitude-based. The phased-based methods are susceptible to noise. The amplitude-based approaches, on the other hand, work well for the scenes that include isolated targets. The entropy-based approach that we propose in this paper, however, will not suffer from the aforementioned shortcomings. 
To verify the accuracy and effectiveness of the proposed approach we utilize experimental data gathered from the RADARSAT-1 satellite \cite{RADARSAT} from Vancouver Canada. To perform the SAR image reconstruction, we exploit the well-known high resolution Range-Doppler algorithm.
The Range-Doppler algorithm is the first algorithm that was developed for spaceborne SAR image reconstruction \cite{Cumming_RD, Chen_RD, Walker_RD}.

The organization of the paper is as follows.
In Section \ref{Model Description}, we present the system model and describe the Range-Doppler algorithm. We have dedicated Section \ref{Doppler-Centroid Estimation} to the Doppler centroid frequency estimation. In Section \ref{Speckle Noise}, we address the speckle noise reduction procedure.
Finally, Section \ref{experimental results} has been dedicated to applying the proposed approaches to the experimental data followed by the concluding remarks. 

\section{Model Description and Image Formation}\label{Model Description}
\begin{figure}
\centerline{
\includegraphics[height=5cm,width=6cm]{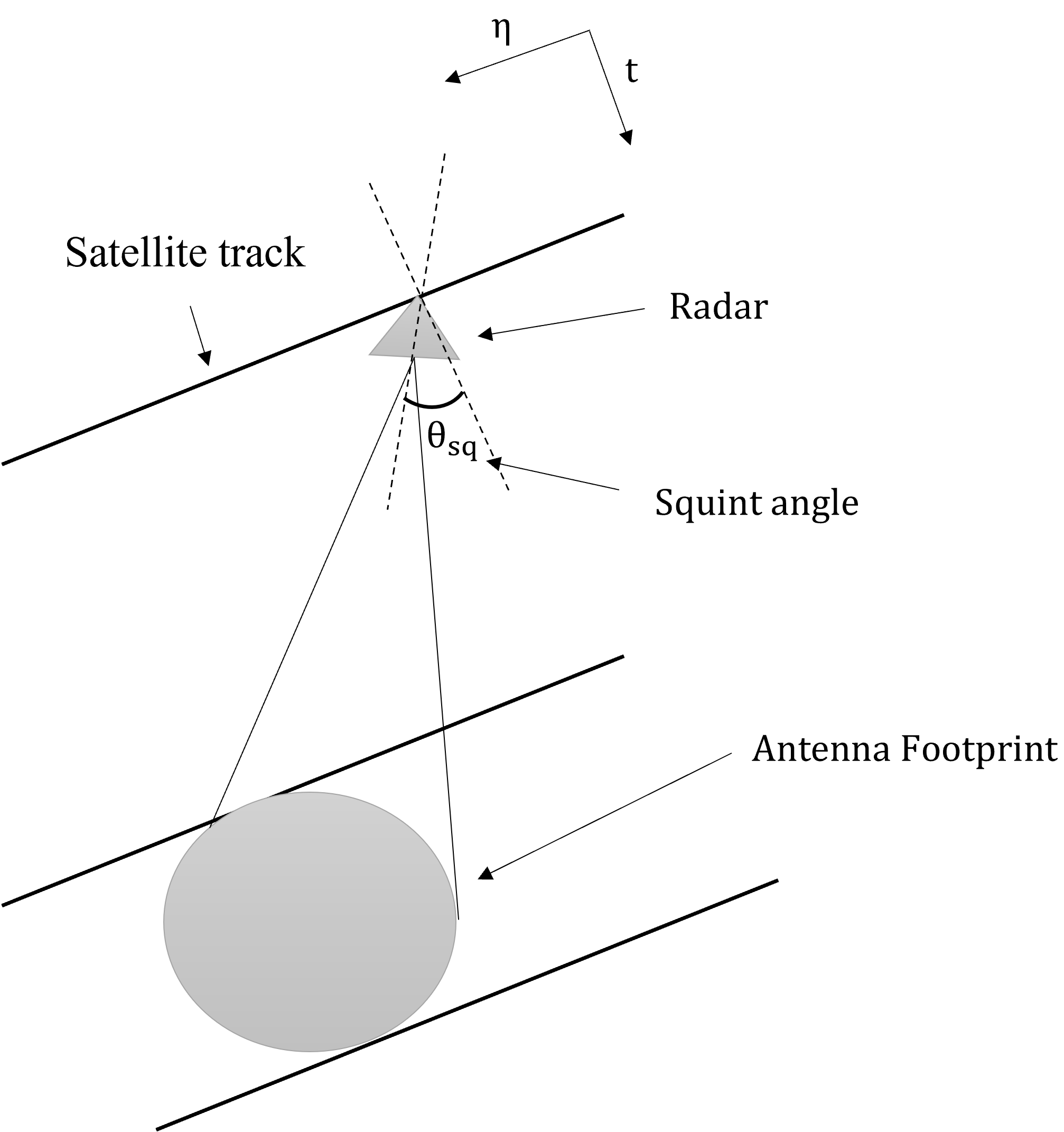}
\hspace{0.1cm}
}
\vspace*{0.1cm}
\caption{The model geometry.
\label{fig:Geometry}}
\end{figure}

Fig.~\ref{fig:Geometry} shows the system model.
The signal transmitted by the radar is a chirp signal which after reflecting back from a point target and after down-conversion is described as 
\begin{dmath}
\label{RX_signal}
s(t, \eta) = \sigma \; w_r(t - \frac{2R(\eta)}{c})w_a(\eta - \eta_c) \times
e^{\displaystyle  -j 4 \pi f_c\frac{R(\eta)}{c} + j \pi \beta t^2 + j 4 \pi \beta t \frac{R(\eta)}{c}}, 
\end{dmath}
where $f_c$ is the carrier frequency and the parameter $\beta$ is given as $b/T$, in which $b$ and $T$ stand for the bandwidth and the chirp time, respectively. In addition, $w_r$ is a rectangular window with length $T$ and $t$ is referred to as fast-time parameter.
In addition, $\sigma$ is the complex reflection coefficient for the point reflector, $\eta$ is referred to as slow-time parameter and $R(\eta)$ is the instantaneous radial distance between the radar and the target which is given as  $R(\eta) = \sqrt{R^2_0 + v^2(\eta - \eta_c)^2}$. Moreover, $w_a$ is a rectangular window with its length equal to the synthetic aperture length divided by $v$, where $v$ is the speed of the satellite.
The ultimate goal in SAR imaging is to estimate the complex reflection  coefficient $\sigma$.

In the following we present the Range-Doppler algorithm \cite{Cumming_RD, Chen_RD, Walker_RD}.
The first step in the Range-Doppler Algorithm is to perform range compression which results in \cite{Cumming}
\begin{dmath}
\label{Range_Compressed}
s_{\rm rc}(t, \eta) = \sigma \; p_r(t - \frac{2R(\eta)}{c})w_a(\eta - \eta_c) 
           e^{\displaystyle  -j 4 \pi f_c \frac{R(\eta)}{c}},
\end{dmath}
where $p_r(x) = \frac{sin(\pi x)}{\pi x}$.
Next, we should compensate for the RCM phenomenon which is the result of coupling between the range and azimuth directions. Consequently, after RCM compensation, we can describe the signal in the Doppler domain as \cite{Cumming}
\begin{dmath}
\label{RCMC}
S_{\rm rcmc}(t, f_\eta) = \sigma \; p_r(t - \frac{2R_0}{c})W_a(f_{\eta} - f_{\eta_c}) \times
           e^{\displaystyle -j \frac{4 \pi f_c R_0}{c}} e^{\displaystyle  j \pi \frac{f^2_{\eta}}{K_a}},
\end{dmath}
in which $K_a = \frac{2v^2}{\lambda R_0}$.
The final step for image formation in the Range-Doppler algorithm, is to compress the data in the azimuth direction which upon performing this stage the compressed signal in the range and azimuth directions is described as \cite{Cumming}
\begin{dmath}
\label{Image_RD}
s_{\rm rcac}(t, \eta) = \sigma \; p_r(t - \frac{2R_0}{c})p_a(\eta) e^{\displaystyle  -j \frac{4 \pi f_c R_0}{c}}
           \;e^{\displaystyle  j 2 \pi f_{\eta_c} \eta},
\end{dmath}
where $f_{\eta_c}$ is the Doppler-Centroid frequency which we study it in the paper extensively.
\section{Doppler-Centroid Estimation}\label{Doppler-Centroid Estimation}
In the case of non-zero squint angle the signal in the azimuth direction is no longer a base-band signal. In other words, the center of the spectrum is not at zero Doppler frequency.
In strip-map mode, the Doppler centroid frequency is the point at which the target is at the center of the antenna main-lobe and hence, it receives the maximum energy from the radar.

Estimating the Doppler centroid frequency can be performed either through the geometry or by using the data. In the case of spaceborne SAR imaging, due to the lack of accuracy in estimating the location of the satellite the Doppler centroid estimation is mainly performed based on the received data.
Since the sampling rate in the azimuth direction is equal to the PRF, therefore, if the Doppler centroid frequency exceeds the PRF, there will be ambiguity in its estimated value.
In fact, we can write the Doppler centroid frequency $f_{\rm dc}$ as
\begin{eqnarray}
\label{Doppler_Centriod_fraction}
f_{\rm dc} &=   M\times {\rm PRF} + f^{\prime}_{\rm dc}  \;\;\; M \in \mathbb{Z},
\end{eqnarray}
where $-\frac{\rm PRF}{2} \leq f^{\prime}_{\rm dc} \leq \frac{\rm PRF}{2}$.
The Doppler centroid frequency depends on the distance between the radar and the target as well as the velocity of the satellite.
The slant range distance from the radar to the target can be Taylor expanded as \cite{Johnson_fdc}
\begin{eqnarray}
\label{R_eta}
R(\eta)  = R(\eta_c) + \frac{dR(\eta)}{d\eta}|_{\eta = \eta_c}  (\eta-\eta_c) \\ \nonumber
                     + \frac{1}{2}\frac{d^2R(\eta)}{d\eta^2}|_{\eta = \eta_c} (\eta-\eta_c)^2.
\end{eqnarray}
Based on (\ref{R_eta}), the Doppler centroid frequency $f_{\rm dc}$ can be expressed as 
\begin{eqnarray}
\label{Doppler_Centriod}
f_{\rm dc}  = \frac{2}{\lambda}\frac{dR(\eta)}{d\eta}|_{\eta = \eta_c} = \frac{2}{\lambda}\frac{v^2\eta_c}{R(\eta_c)}.
\end{eqnarray}
The third term in (\ref{R_eta}) is called the FM rate in slow time. 
From (\ref{R_eta}), it is obvious that, the signal in the azimuth direction is a non-base-band chirp signal. 
The fact that the relative motion between the satellite and the target creates a chirp signal is the reason for achieving high resolution in the azimuth direction.

There are different techniques for the Doppler centroid frequency estimation and they are divided into two different categories, namely, amplitude-based and phased-based methods \cite{Cumming}.
The amplitude-based spectral fit algorithm and phase-based Average Cross Correlation Coefficient (ACCC) method are among two well-known methods for estimating $f^{\prime}_{\rm dc}$ \cite{Cumming, Madsen}.
The fractional part of the Doppler centroid frequency is required to localize the energy of the targets in the azimuth direction. However, to compensate for the effect of the RCM, we require the unwrapped Doppler centroid frequency.
The phased-based methods are sensitive to the noise and the amplitude-based techniques work well when there are isolated targets with strong reflection coefficients. Therefore, we propose the entropy-based approach as 
 \begin{eqnarray}
\label{Entropy}
\mathcal{I}  =  - \sum_{i=1}^{n} \sum_{j=1}^{m} |\rm img[i,j]|\log_2\left(|\rm img[i,j]|\right),
\end{eqnarray}
where $|\rm img[i,j]|$ represents the normalized absolute value for the $(\rm ij)^{\rm th}$ pixel of the reconstructed image. Furthermore, $m$ and $n$ are the number of pixels in the range and along-track directions, respectively. Moreover, $\mathcal{I}$ is the entropy of the image. When the image is focused, the entropy of the image is at its minimum level. 

Multilook Cross Correlation (MLCC), Wavelength Diversity Algorithm (WDA) \cite{Bamler_fdc} and Multilook Beat Frequency (MLBF) technique \cite{Shu_fdc_MLBF_2, Shu_fdc_MLBF, Cumming} are among the methods that are used for estimating the unambiguous value of the Doppler centroid frequency \cite{Cumming}. These three techniques are all phased-based methods.
In the case of having strong isolated targets in the scene, the unambiguous Doppler centroid frequency can be calculated by estimating the slope of the range compressed data for the isolated targets. In fact, this is the technique we exploit to estimate the unambiguous Doppler centroid. 
\section{Speckle Noise}\label{Speckle Noise}
When the roughness of the terrain to be imaged is on the scale of the wavelength of the incident wave, speckle noise appears. Speckle noise is a multiplicative noise.
In fact, the origin of the speckle noise is the dependency of the relative phase of the individual reflectors inside the resolution cell on the viewing angle \cite{Cumming}.
The well-known method for speckle noise reduction is multi-look processing in which the synthetic aperture is divided into several independent sections and per each section an image from the scene is created \cite{Cumming}. The final image is the summation of the absolute value of the images created from the sub-apertures. In other words, non-coherent averaging of the reconstructed images yields the final image. 
The major problem with the multi-look processing is that the resolution reduces due to using only a small portion of the aperture when we create each independent images from the scene.
In this section, we utilize median filtering approach over the reconstructed image to reduce the effect of the speckle noise. We use the entire synthetic aperture to reconstruct the image, therefore, there is no loss in the resolution of the reconstructed image.
We introduce a 2D $m \times n$ filter and slide it over the reconstructed image while solving the following optimization problem,
\begin{eqnarray}
\label{speckle}
\!\min_{a}        &\qquad&    \sum_{i=1}^{n} \sum_{j=1}^{m}| a_{\rm ij}-a|,
\end{eqnarray}
where $a_{\rm ij}$ is the value for the $(\rm ij)^{\rm th}$ pixel and $a$ is the value selected by the optimization problem for the $(\lfloor \frac{n-1}{2} \rfloor + 1 \; \lfloor \frac{m-1}{2} \rfloor + 1)^{\rm th}$ pixel. In fact, by solving the optimization problem in (\ref{speckle}), we are filtering the image by the median filter. In other words, we replace the value of each pixel with the median of the neighboring pixels. This is proved to be a powerful method to decrease the effect of the speckle noise which, as a result, the fine structures of the image can be revealed.

\section{Experimental Results}\label{experimental results}
\begin{figure}
\centering
\begin{tikzpicture}[yshift=0.00001cm][font=\small]
  \node (img1)  {\includegraphics[height=5cm,width=7cm]{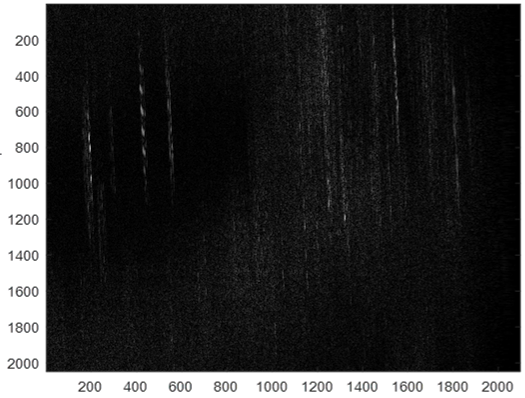}};
  \node[left=of img1, node distance=0cm, rotate = 90, xshift=1.5cm, yshift=-0.9cm,font=\color{black}] {{Along-Track [Samples]}};
  \node[below=of img1, node distance=0cm, xshift=0cm, yshift=1.1cm,font=\color{black}] {{Slant-Range [Samples]}};
\node[below=of img1, node distance=0cm, xshift=0cm, yshift=0.7cm,font=\color{black}, font = \small] {{(a)}};
\end{tikzpicture}
\begin{tikzpicture}[yshift=0.00001cm][font=\small]
\node(img2) {\includegraphics[height=5cm,width=7cm]{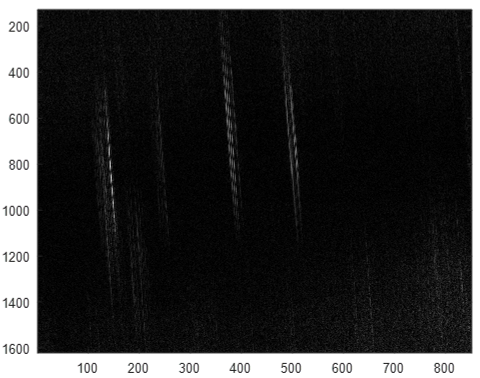}};
  \node[left=of img1, node distance=0cm, rotate = 90, xshift=1.5cm, yshift=-0.9cm,font=\color{black}] {{Along-Track [Samples]}};
  \node[below=of img1, node distance=0cm, xshift=0cm, yshift=1.1cm,font=\color{black}] {{Slant-Range [Samples]}};
 \node[below=of img2, node distance=0cm, xshift=0cm, yshift=0.7cm,font=\color{black}, font = \small] {{(b)}};

\end{tikzpicture}
\caption{a) the result of applying (\ref{Range_Compressed}) to the raw data and obtaining the range compressed signal, b) the blow-up part of Fig.~\ref{fig:Range_Compressed}-(a) which shows the range compressed result for several strong isolated targets.}
\label{fig:Range_Compressed}
\end{figure}

In this section, we present the result of SAR image reconstruction based on the experimental data gathered from the Canadian RADARSAT-1 satellite. The data is from Vancouver Canada. The total data collection has taken 15 seconds.
The specifications of the  RADARSAT-1 satellite have been given in Table.\ref{Tab:Table}.
\begin{table}
  \centering
  \caption{RADARSAT-1's Parameters}\label{Tab:Table}
  \begin{center}
    \begin{tabular}{| l | l | l |}

    \hline
    Parameters &  & Values  \\ \hline
    Center frequency ($\rm GHz$) & $\rm f_c$ & $5.3$ \\ \hline
    Radar sampling rate ($\rm MHz$) & $\rm Fr$ & $32.317$  \\ \hline
    Pulse repetition frequency ($\rm Hz$) & $\rm PRF$ & $1256.98$ \\ \hline
    Slant range of first radar sample ($\rm km$) & $R_0$ & $988.65$  \\ \hline
    FM rate of radar pulse (${\rm MHz}/\mu s$) & $ \beta$     &  $0.72135$   \\ \hline
    Chirp duration ($ \mu s$)  &  $\rm T$  &  $41.75$   \\ \hline
    Satellite velocity ($\rm m/s$) &   $\rm v$  &  $7062$    \\ \hline
    Bandwidth ($\rm MHz$)  &  $\rm b$   & $30.116$ \\ \hline
    \hline
    \end{tabular}
\end{center}
\end{table}
We select the data related to English Bay. The reason is, there are several ships in this scene that play the role of isolated strong reflectors which will help to see the effect of the RCM clearly and will also be utilized to perform Doppler centroid estimation. 
We apply the Range-Doppler algorithm to the raw data. Fig.~\ref{fig:Range_Compressed}-(a) shows the range compressed data based on (\ref{Range_Compressed}).
The range compressed energy of the ships can be seen as a few skewed vertical lines. The skew demonstrates the effect of the RCM.

\begin{figure}
\begin{tikzpicture}[yshift=0.00001cm][font=\small]
  \node (img1)  {\includegraphics[height=3cm,width=3cm]{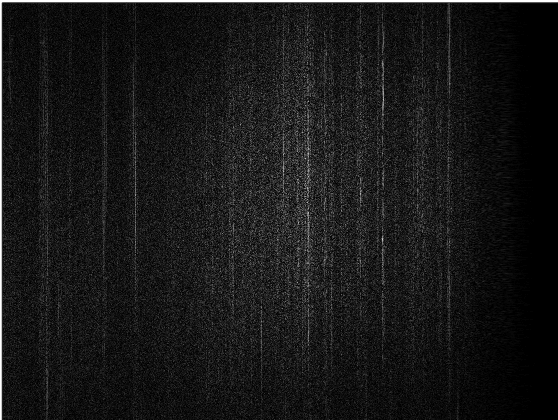}};
  \node[left=of img1, node distance=0cm, rotate = 90, xshift=1cm, yshift=-0.9cm,font=\color{black}] {{Along-Track }};
  \node[below=of img1, node distance=0cm, xshift=0cm, yshift=1.1cm,font=\color{black}] {{Slant-Range}};
\node[below=of img1, node distance=0cm, xshift=0cm, yshift=0.8cm,font=\color{black}] {{(a)}};
\hspace{3.8cm}
\node(img2) {\includegraphics[height=2.7cm,width=4cm]{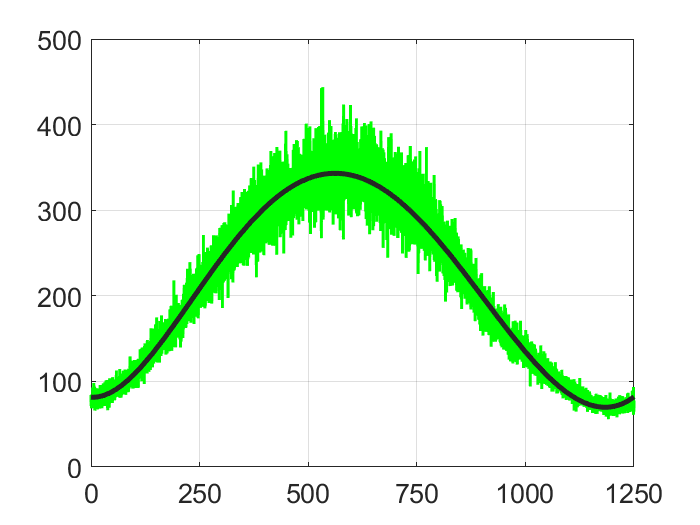}};
  \node[left=of img1, node distance=0cm, rotate = 90, xshift=0.8cm, yshift=-0.6cm,font=\color{black}] {{$\rm Intensity$}};
  \node[below=of img1, node distance=0cm, xshift=0cm, yshift=1.3cm,font=\color{black}] {{Frequency [Hz]}};
 \node[below=of img2, node distance=0cm, xshift=0cm, yshift=0.8cm,font=\color{black}] {{(b)}};

\end{tikzpicture}
\caption{a) the range compressed image after RCM compensation based on (\ref{RCMC}), b) the amplitude-based estimation of the fractional part of the Doppler centroid frequency with its maximum value at $f^{\prime}_{\rm dc} = 520 \; \rm Hz$.}
\label{fig:RCMC}
\end{figure}

In order to perform the RCM compensation we are required to estimate the unambiguous Doppler centroid frequency.
As we mentioned before, if the Doppler centroid frequency is larger than the PRF of the RADAR, there will be an ambiguity in the Doppler centroid frequency estimation.
One way to estimate the unambiguous value for the Doppler centroid frequency is by analyzing the trajectory of the strong isolated targets.
Fig.~\ref{fig:Range_Compressed}-(b) shows a blow-up part of the range compressed image shown in Fig.~\ref{fig:Range_Compressed}-(a) related to a few stationary ships in water.
The skew in the trajectory of these targets is due to the non-zero Doppler centroid frequency. The energy of each of these targets is spread over several different range cells.
The slope can easily be calculated and is equal to $0.034$ range samples per azimuth samples. To estimate the Doppler centroid frequency we should first multiply the slope by $\rm \frac{c}{2Fr}$ to convert the slope from range samples to range distance and then multiply it by $\rm PRF$ to convert it from azimuth samples to azimuth time.
Hence, we have $\frac{dR(\eta)}{d\eta} = 198.23 \; \rm m/s$. As a result, from (\ref{Doppler_Centriod}) the Doppler centroid frequency can be calculated as $f_{\rm dc} = -7009 \; \rm Hz$.
Fig.~\ref{fig:RCMC}-(a) illustrates the result of the RCM compensation process. As can be seen from Fig.~\ref{fig:RCMC}-(a), the energy of the targets have been localized in their corresponding range cells.

\begin{figure}
\centering
\includegraphics[height=6cm,width=8cm]{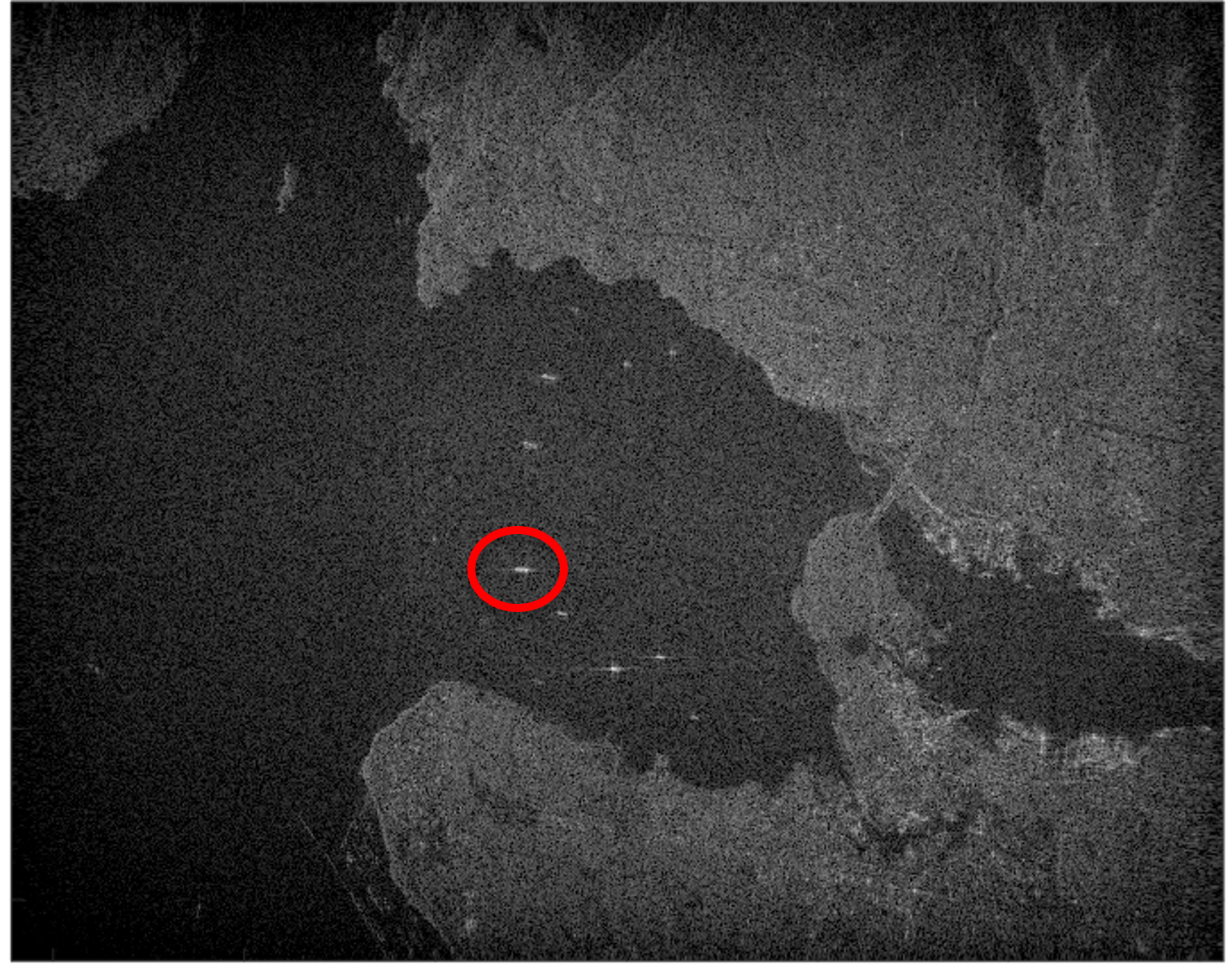}
\caption{{The reconstructed image from English Bay based on the Range-Doppler algorithm}.
\label{fig:Image_Bay}}
\end{figure}

Next, we estimate the fractional part of the Doppler centroid frequency which is $f^{\prime}_{\rm dc}$. The fractional part is essential in focusing the energy of the targets in the azimuth direction. 
Fig.~\ref{fig:RCMC}-(b) illustrates the power spectrum of the data in azimuth direction versus slow time frequency $f_{\eta}$. In order to reduce the effect of the noise, we have averaged the power spectrum corresponding to $230$ range cells. In order to be able to calculate the frequency component at which the signal reaches its maximum value, we have performed curve fitting.
From Fig.~\ref{fig:RCMC}-(b), we can easily estimate the fractional part of the Doppler centroid frequency as $f^{\prime}_{\rm dc} = 520\; \rm Hz$.
Finally, at the last stage, we perform the azimuth localization based on (\ref{Image_RD}). Consequently, the reconstructed image based on the Range-Doppler algorithm is obtained which has been presented in Fig.~\ref{fig:Image_Bay}.

\begin{figure}
\begin{tikzpicture}[yshift=0.00001cm][font=\small]
  \node (img1)  {\includegraphics[height=2.5cm,width=3cm]{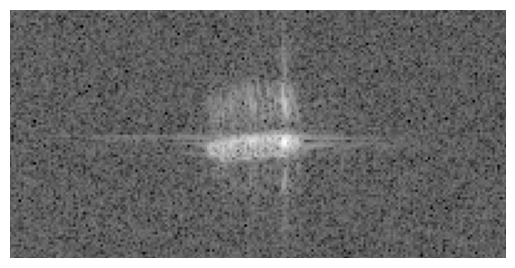}};
  \node[left=of img1, node distance=0cm, rotate = 90, xshift=1cm, yshift=-0.9cm,font=\color{black}] {{Along-Track}};
  \node[below=of img1, node distance=0cm, xshift=0cm, yshift=1.1cm,font=\color{black}] {{Slant-Range}};
\node[below=of img1, node distance=0cm, xshift=0cm, yshift=0.8cm,font=\color{black}] {{(a)}};
\hspace{4.5cm}
\node(img2) {\includegraphics[height=2.5cm,width=3cm]{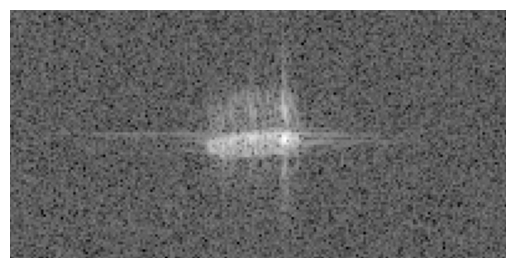}};
  \node[left=of img1, node distance=0cm, rotate = 90, xshift=1cm, yshift=-0.9cm,font=\color{black}] {{Along-Track}};
  \node[below=of img1, node distance=0cm, xshift=0cm, yshift=1.1cm,font=\color{black}] {{Slant-Range}};
\node[below=of img1, node distance=0cm, xshift=0cm, yshift=0.8cm,font=\color{black}] {{(b)}};
\end{tikzpicture}

\begin{tikzpicture}[yshift=0.00001cm][font=\small]
  \node (img1)  {\includegraphics[height=2.5cm,width=3cm]{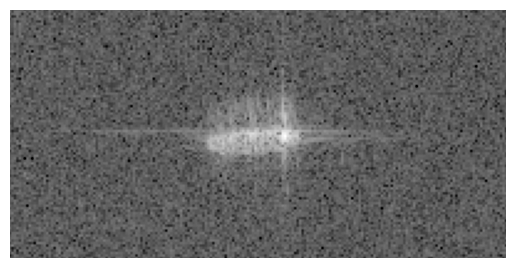}};
   \node[left=of img1, node distance=0cm, rotate = 90, xshift=1cm, yshift=-0.9cm,font=\color{black}] {{Along-Track}};
  \node[below=of img1, node distance=0cm, xshift=0cm, yshift=1.1cm,font=\color{black}] {{Slant-Range}};
\node[below=of img1, node distance=0cm, xshift=0cm, yshift=0.8cm,font=\color{black}] {{(c)}};
\hspace{4.5cm}
\node(img2) {\includegraphics[height=2.5cm,width=3cm]{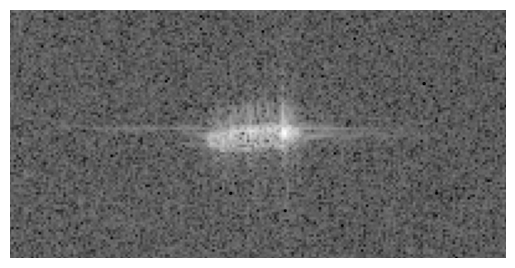}};
   \node[left=of img1, node distance=0cm, rotate = 90, xshift=1cm, yshift=-0.9cm,font=\color{black}] {{Along-Track}};
  \node[below=of img1, node distance=0cm, xshift=0cm, yshift=1.1cm,font=\color{black}] {{Slant-Range}};
\node[below=of img1, node distance=0cm, xshift=0cm, yshift=0.8cm,font=\color{black}] {{(d)}};
\end{tikzpicture}

\begin{tikzpicture}[yshift=0.00001cm][font=\small]
\centering
  \node (img1)  {\includegraphics[height=2.5cm,width=3cm]{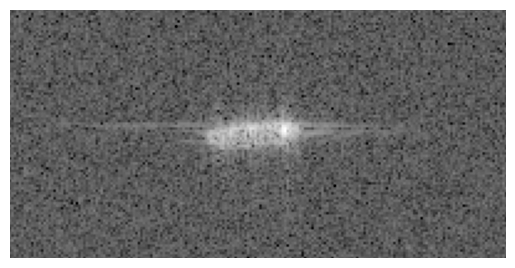}};
   \node[left=of img1, node distance=0cm, rotate = 90, xshift=1cm, yshift=-0.9cm,font=\color{black}] {{Along-Track}};
  \node[below=of img1, node distance=0cm, xshift=0cm, yshift=1.1cm,font=\color{black}] {{Slant-Range}};
\node[below=of img1, node distance=0cm, xshift=0cm, yshift=0.8cm,font=\color{black}] {{(e)}};

\hspace{4.5cm}
\node(img2) {\includegraphics[height=2.5cm,width=3cm]{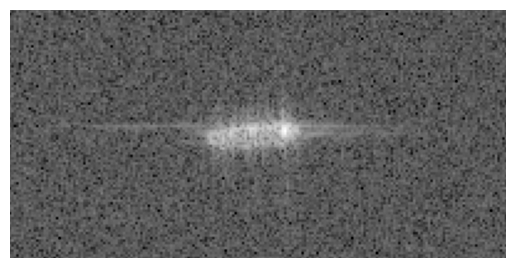}};
   \node[left=of img1, node distance=0cm, rotate = 90, xshift=1cm, yshift=-0.9cm,font=\color{black}] {{Along-Track}};
  \node[below=of img1, node distance=0cm, xshift=0cm, yshift=1.1cm,font=\color{black}] {{Slant-Range}};
\node[below=of img1, node distance=0cm, xshift=0cm, yshift=0.8cm,font=\color{black}] {{(f)}};
\end{tikzpicture}

\caption{The reconstructed images of one of the ships, marked with red circle in Fig.~\ref{fig:Image_Bay}, based on $f^{\prime}_{\rm dc}$ equal to, a) $-600 \; \rm Hz$, b) $-300 \; \rm Hz$, c) $0 \; \rm Hz$, d) $300 \; \rm Hz$, e) $516 \; \rm Hz$, f) $600 \; \rm Hz$.}
\label{fig:Entropy_image}
\end{figure}
\begin{figure}
\centering
\begin{tikzpicture}[yshift=0.00001cm][font=\large]
\node(img2) {\includegraphics[height=5.5cm,width=7.5cm]{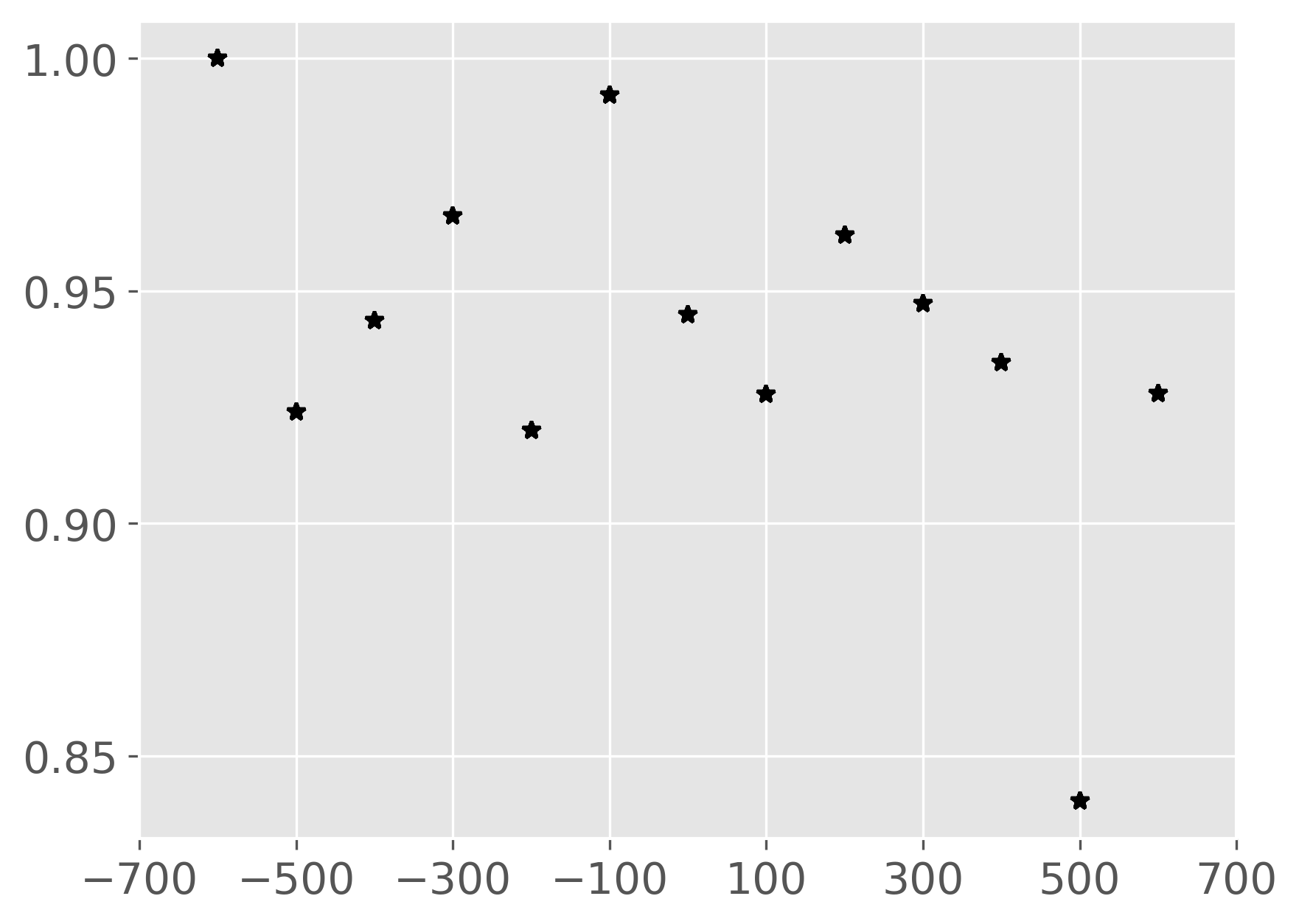}};
  \node[left=of img1, node distance=0cm, rotate = 90, xshift=2.2cm, yshift=1.2cm,font=\color{black}] {{Normalized Entropy}};
  \node[below=of img1, node distance=0cm, xshift=0.4cm, yshift=-0.3cm,font=\color{black}] {{Frequency [Hz]}};
\node[below=of img1, node distance=0cm, xshift=0.2cm, yshift=-0.7cm,font=\color{black}, font = \small] {{(a)}};

\end{tikzpicture}

\begin{tikzpicture}[yshift=0.00001cm][font=\large]
\node(img2) {\includegraphics[height=5.5cm,width=7.5cm]{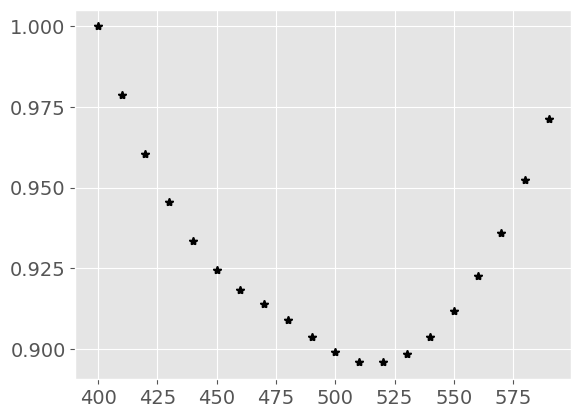}};
  \node[left=of img1, node distance=0cm, rotate = 90, xshift=2.2cm, yshift=1.2cm,font=\color{black}] {{Normalized Entropy}};
  \node[below=of img1, node distance=0cm, xshift=0.4cm, yshift=-0.3cm,font=\color{black}] {{Frequency [Hz]}};
\node[below=of img1, node distance=0cm, xshift=0.2cm, yshift=-0.7cm,font=\color{black}, font = \small] {{(b)}};

\hspace{0.9 cm}
\vspace{2 cm}
\node(img2) {\includegraphics[height=2cm,width=2.5cm]{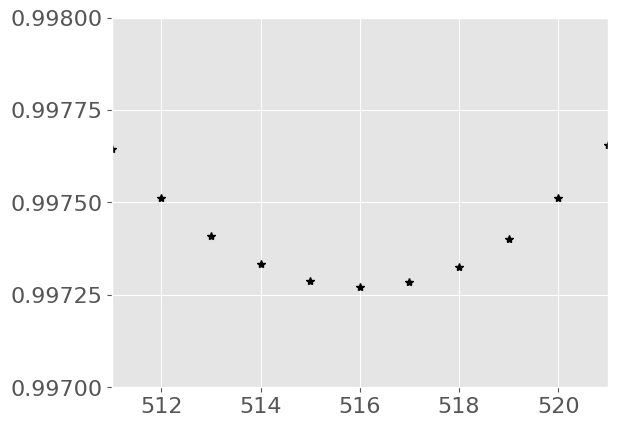}};
  \node[left=of img1, node distance=0cm, rotate = 90, xshift=1cm, yshift=-1.2cm,font=\color{black}, font = \tiny] {{Normalized Entropy}};
  \node[below=of img1, node distance=0cm, xshift=0cm, yshift=1.4cm,font=\color{black},  font = \tiny] {{Frequency [Hz]}};

\end{tikzpicture}
\caption{The entropy-based estimation of the fractional part of the Doppler centroid frequency for different values for $f^{\prime}_{\rm dc}$ ranging from a) $-600 \; \rm Hz$ to  $600 \; \rm Hz$ with the step-size equal to $100 \; \rm Hz$, b) $400 \; \rm Hz$ to  $600 \; \rm Hz$ with the step-size equal to $10 \; \rm Hz$ as well as the magnified part of the graph for $512 \; \rm Hz$ to  $520 \; \rm Hz$ with the step-size equal to $1 \; \rm Hz$.}
\label{fig:Entropy}
\end{figure}

In order to present the entropy-based approach, we select an isolated target with strong reflection coefficient in Fig.~\ref{fig:Image_Bay}. In fact, we have selected one of the ships inside the red circle in Fig.~\ref{fig:Image_Bay} and have utilized the data from this part of the scene to calculate the fractional part of the Doppler centroid frequency.
Fig.~\ref{fig:Entropy_image} illustrates the result of reconstructing the image based on different values for $f^{\prime}_{\rm dc}$.  From Fig.~\ref{fig:Entropy_image}, it is obvious that when we deviate from $520 \; \rm Hz$, image de-focusing is occurred. 
To estimate the fractional part of the Doppler centroid using the entropy-based approach, we calculate the entropy of the reconstructed images using (\ref{Entropy}) and based on different values for $f^{\prime}_{\rm dc}$ ranging from $-600 \; \rm Hz$ to  $600 \; \rm Hz$ with the step-size equal to $100 \; \rm Hz$. The result has been presented in Fig.~\ref{fig:Entropy}-(a).  From Fig.~\ref{fig:Entropy}-(a), it is clear that the minimum entropy occurs in close proximity of $500 \; \rm Hz$. Subsequently, to obtain a more accurate result, we have searched through the range of frequencies from  $400 \; \rm Hz$ to  $600 \; \rm Hz$ and have shown the result in Fig.~\ref{fig:Entropy}-(b). As it is clear from Fig.~\ref{fig:Entropy}-(b) and specifically from the magnified part of the graph, the minimum value for the entropy is achieved when $f^{\prime}_{\rm dc} = 516 \; \rm Hz$. In fact, $f^{\prime}_{\rm dc} = 516 \; \rm Hz$ is the estimated value that we have obtained using the proposed entropy-based method. 

\begin{figure}
\centering
\includegraphics[height=6cm,width=8cm]{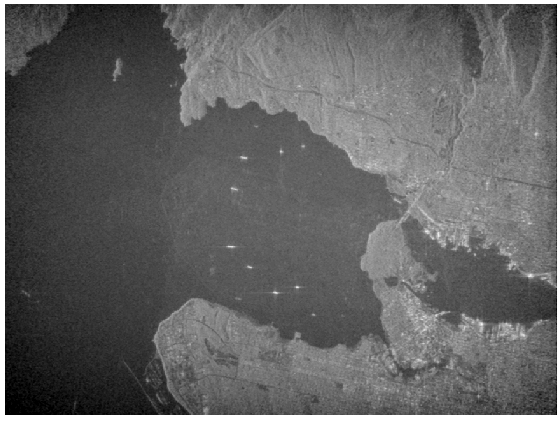}
\caption{{The result of the speckle noise reduction process, based on (\ref{speckle}), for the reconstructed image of English Bay shown in Fig.~\ref{fig:Image_Bay}}.
\label{fig:EnglishBay_Speckle}}
\end{figure}
\begin{figure}
\includegraphics[height=6cm,width=8cm]{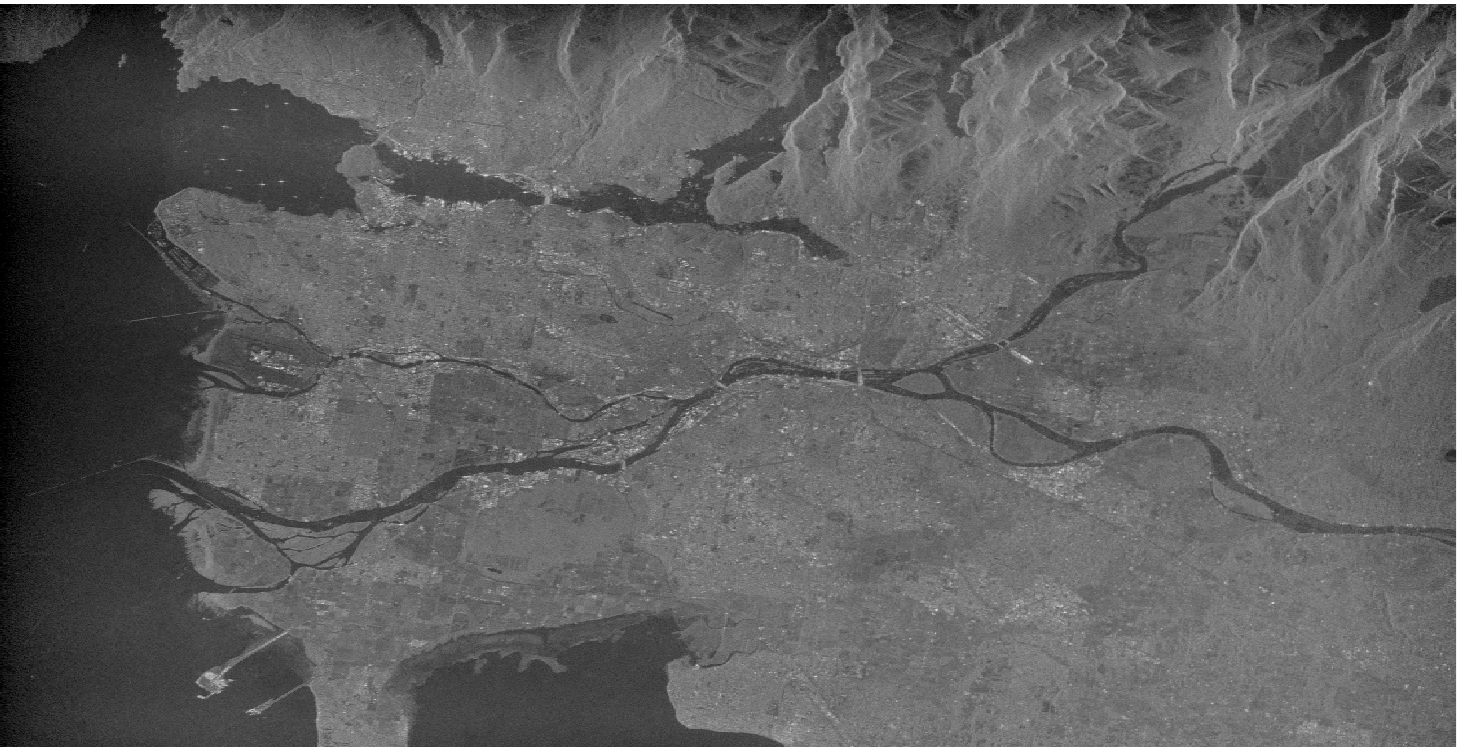}
\centering
\caption{{The reconstructed image from Vancouver, Canada based on the Range-Doppler algorithm}.
\label{fig:RD_Image}}
\end{figure}

After the image reconstruction is completed, we will then focus on the speckle noise removal procedure based on (\ref{speckle}). 
Fig.~\ref{fig:EnglishBay_Speckle} shows the result of speckle noise reduction for the reconstructed image depicted in Fig.~\ref{fig:Image_Bay}. To perform the speckle noise reduction, we have chosen $m=n=6$ for the filter given in (\ref{speckle}).
Fig.~\ref{fig:RD_Image} shows the result of applying the Range-Doppler algorithm to a larger part of the raw data which shows the entire Vancouver area. To remove the effect of the speckle noise we have applied the filter given in (\ref{speckle}) with $m=n=10$.

\section{Conclusion}\label{conclusion}
In this paper, we presented the sapceborne SAR imaging process based on the well-known high resolution Range-Doppler algorithm. We, then, focused on the fractional part of the Doppler centroid and proposed the entropy-based method to estimate the parameter.
At the end, we verified the accuracy of the proposed approach by utilizing experimental data. We further addressed the speckle noise removal process and presented the results.

As future work, we will focus on comparing the existing methods with the proposed technique for the Doppler centroid frequency estimation and measure the improvement with numeric metrics.

\bibliographystyle{IEEEtran}
\bibliography{Biblio}

\begin{IEEEbiography}[{\includegraphics[width=1in,height=1.25in,clip,keepaspectratio]{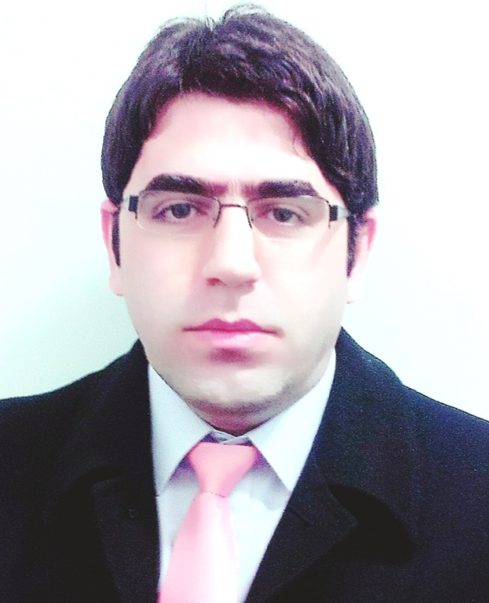}}]{Shahrokh Hamidi} was born in 1983, in Iran. He received his B.Sc., M.Sc., and Ph.D. degrees all in Electrical and Computer Engineering. He is with the Faculty of Electrical and Computer Engineering at the University of Waterloo, Waterloo, Ontario, Canada. His current research areas include statistical signal processing, mmWave and Terahertz imaging, image processing, system design,  multi-target tracking, wireless communication, machine learning, and optimization.
\end{IEEEbiography}

\end{document}